\documentclass[iop,english,twocolappendix,numberedappendix,showpacs,superscriptaddress,appendixfloats,tighten,apj,twocolumn]{aastex6}
\usepackage{amsmath,amsfonts,amssymb,xcolor,soul}
\usepackage{dcolumn,ulem}
\usepackage{bm}
\usepackage{float}
\usepackage{hyperref}

\usepackage{amstext}
\usepackage{graphicx}
\usepackage{babel}
\usepackage{color}

\def\be{\begin{equation}}
\def\ee{\end{equation}}
\def\ba{\begin{eqnarray}}
\def\ea{\end{eqnarray}}

\def \pmbtext#1{\leavevmode
     \setbox0\hbox{#1}
     \kern0,4pt \copy0 \kern-\wd0
     \kern-0,2pt \raise0,3pt \box0 }

\newcommand{\el}{\boldsymbol{\ell}}
\newcommand{\va}{\textbf{b}}

\newcommand{\vh}{\textbf{v}}

\newcommand{\ja}{\textbf{j}}

\newcommand{\lang}{\left\langle}
\newcommand{\rang}{\right\rangle}

\newcommand{\nab}{\boldsymbol\nabla}

\usepackage{soul}

\begin{document} 

\title{On exact laws in incompressible Hall magnetohydrodynamic turbulence}

\author{R. Ferrand\altaffilmark{1}, S. Galtier\altaffilmark{1,2}, F. Sahraoui\altaffilmark{1},  R. Meyrand\altaffilmark{1}, N. Andr\'es\altaffilmark{1,3} and S. Banerjee\altaffilmark{4}} 
\email{renaud.ferrand@lpp.polytechnique.fr}
\affil{$^1$ LPP/CNRS-Ecole polytechnique-Sorbonne Universit\'e-Universit\'e Paris-Sud-Observatoire de Paris, 91128 Palaiseau, France}
\affil{$^2$ Institut Universitaire de France}
\affil{$^3$ Instituto de Astronom\'ia y F\'isica del Espacio, UBA-CONICET, CC. 67, suc. 28, 1428, Buenos Aires, Argentina.}
\affil{$^4$ Department of Physics, Indian Institute of Technology Kanpur, Kalyanpur 208016, Uttar Pradesh, India}


\begin{abstract}
A comparison is made between several existing exact laws in incompressible Hall magnetohydrodynamic (IHMHD) turbulence in order to show their equivalence, despite stemming from different mathematical derivations. Using statistical homogeneity, we revisit the law proposed by Hellinger et al. (2018) and show that it can be written, after being corrected by a multiplicative factor, in a more compact form implying only flux terms expressed as increments of the turbulent fields. The Hall contribution of this law is tested and compared to other exact laws derived by Galtier (2008) and Banerjee \& Galtier (2017) using direct numerical simulations (DNSs) of three-dimensional electron MHD (EMHD) turbulence with a moderate mean magnetic field. 
We show that the studied laws are equivalent in the inertial range, thereby offering several choices on the formulation to use depending on the needs. The expressions that depend explicitly on a mean (guide) field may lead to residual errors in estimating the energy cascade rate ; however, we demonstrate that this guide field can be removed from these laws after mathematical manipulation. Therefore, it is recommended to use an expression independent of the mean guide field to analyze numerical or {\it in-situ} spacecraft data.
\end{abstract}

\maketitle

\section{Introduction}

To date, understanding the dynamics of turbulent flows remains one of the most challenging problems of classical physics. As these systems are inherently chaotic they are generally studied by statistical means, thus requiring specific tools to be dealt with \citep{frisch}. The so-called exact laws are among the most important theoretical results of turbulence. The derivation of these statistical laws is based on the assumption of the existence of an inertial range where the physics is dominated by the nonlinear transfer from one scale to another. In a fully developed three dimensional (3D) hydrodynamic turbulence of an incompressible fluid, kinetic energy is transferred from macroscopic length scales to the scale of molecular diffusion until it is eventually dissipated into thermal energy by viscous effects. The mean transfer rate of kinetic energy per unit volume, which is usually denoted by $\varepsilon$, is assumed to remain constant at each scale in the inertial range where both dissipation and forcing mechanisms are negligible. It is also equal to the average energy dissipation rate, which is expected to be independent of the viscosity in the limit of large Reynolds numbers. This property, often called the zeroth law of turbulence \citep{Onsager49,Eyink94,DR00,SaintMichel14}, is actually used to link the fluctuations of the velocity field to $\varepsilon$ through exact laws. 

The first and the most popular exact law is the so-called Kolmogorov's four-fifth law which was derived for incompressible turbulence \citep{kolmogorov}. It was first derived using tensorial calculus \citep{batch}, but a similar four-third law was computed more directly through the dynamical study of an appropriate two-point correlation function~\citep{monin57,antonia97}. Using these methods (with the generalized zeroth law of turbulence \citep{Mininni09,Bandy18,Galtier2018}), new laws were derived for different plasma models such as incompressible MHD (IMHD) \citep{PP98} or IHMHD \citep{galtier08}. More recently, these results were extended to compressible (isothermal and polytropic) turbulence in hydrodynamics (CHD) \citep{GB11,BG2014}, and then to isothermal compressible MHD (CMHD) \citep{Banerjee13,A2017b} and compressible HMHD (CHMHD) \citep{andres18}. 
Using an alternative formulation~\citep{banerjee16a,banerjee17}, compressible exact relations were also derived for self-gravitating turbulence of both neutral and MHD fluids~\citep{BK17,BK18}.
Such laws were also derived for self-gravitating turbulence whose potential applications are the interstellar medium and star formation. 

Since $\varepsilon$ can be used as a proxy to evaluate the amount of energy available to be ultimately dissipated at small scales, exact laws are often used in collisionnless astrophysical plasmas, such as the solar wind (SW), to evaluate the rate of plasma heating. Indeed, \cite{Richardson95} have evidenced using Voyager data a slower decay of the (ion) temperature with the heliocentric distance in comparison with the prediction from the adiabatic expansion model \citep{Matthaeus99}. Turbulence is proposed to explain this problem because it provides an efficient mechanism of energy dissipation through the nonlinear process of energy cascade from the MHD scales down to the sub-ion and electron scales, where the energy is eventually dissipated through some kinetic effects \citep{He15,Sahraoui09,Sahraoui10}. The energy cascade (or dissipation) rate was measured in both the SW \citep{Podesta07,sorriso07,macbride08,Marino08,Carbone09,smith09,Stawarz,Osman,coburn,BanerjeeSW16,hadid17} and the Earth's magnetosheath \citep{hadid18}, and was shown to enlighten many aspects related to the dynamics of turbulent space plasmas. 

Similarly to hydrodynamics, the case of IHMHD has driven some attention over the last ten years with the derivation of several exact laws \citep{galtier08,banerjee17,hellinger18}. Although the underlying assumptions remain unchanged, these laws stem from the analysis of different statistical quantities. On the one hand, the laws given in \cite{galtier08} (hereafter G08) and \cite{banerjee17} (hereafter BG17) are derived from the dynamical analysis of the two-point correlator,
\begin{align} \label{correlator}
       \langle R_E \rangle =& \left\langle \frac{\vh\cdot\vh' + \va\cdot\va'}{2} \right\rangle \, ,
\end{align}
where $\langle \rangle$ is the ensemble average, 
$\vh$ and $\va$ are the local velocity and Alfv\'en velocity fields, respectively, and the prime distinguishes values taken at points ${\bf x}$ and ${\bf x'}$, respectively (see Section \ref{calculation} for the definitions). However, the calculation was done differently in the two models (G08 and BG17) and yielded quite different expressions that cannot be trivially connected to each other. 
On the other hand, the law from Hellinger et al. 2018 \citep{hellinger18} (hereafter H18) stems from the evolution equation of the second order structure function
\begin{align} \label{correlator2}
        \langle S \rangle =& \langle | \vh' - \vh|^2+| \va' - \va |^2 \rangle \, ,
\end{align}
which is linked to expression (\ref{correlator}) through the relation $\lang S/4 \rang = \lang E^{tot} \rang - \lang R_E \rang$ with $E^{tot}=  (\vh^{2} + \va^{2})/2$ the total energy.
It is thus important to check whether or not these different laws are consistent with each other by providing the same energy cascade rate. Note that in the definition (\ref{correlator2}), $\langle S \rangle$ is independent of the (constant) mean fields $\vh_0$ and $\va_0$. We will return to this point in Section~\ref{num}.
    
This paper aims at studying analytically and numerically the IHMHD exact laws and check if they are mathematically equivalent despite stemming from a different logic of derivation. Following this goal, we expose in Section \ref{calculation} a rigorous derivation of H18 and find a slight difference with the original paper. Furthermore, we provide a new more compact form of that law that depends only on flux terms (hereafter F19). In Section \ref{equiv} we give a mathematical proof of the equivalence of the three laws F19, G08 and BG17. This equivalence is eventually tested in Section \ref{num} with 3D DNSs of EMHD turbulence. We also discuss the possible influence of a mean magnetic field on the exact laws and on the methods used to evaluate the energy cascade rate. The results are summarized and discussed in Section \ref{conclusion}. 

\section{Derivation of H18} \label{calculation}

In this section we propose a step-by-step derivation of the H18 law based on the same premises as in the original paper, where the details were not given. We note \textbf{B} the magnetic field and $\textbf{J}= \boldsymbol{\nabla} \times \textbf{B}/\mu_0$ the electric current; 
the mass density $\rho_0$ is taken constant and equal to unity. We use the Alfv\'en units for the magnetic field and the electric current, i.e. $\va = \textbf{B}/\sqrt{\mu_0\rho_0}$ and $\ja = \boldsymbol{\nabla} \times \va$. In the incompressible case (i.e., $\nab\cdot\vh=0$) we get the following velocity and induction equations,
    \begin{align} 
        \partial_t \vh =& -(\vh\cdot\nab)\vh + (\va\cdot\nab)\va - \nab P + {\bf d}_\nu + {\bf f} \label{equations0} \, , \\
        \partial_t \va =& -(\vh\cdot\nab)\va + (\va\cdot\nab)\vh  \nonumber \\
        &+ d_i(\ja\cdot\nab)\va - d_i(\va\cdot\nab)\ja  + {\bf d}_\eta \, , \label{equations} \\
        \nab \cdot \va =& 0 \, , \label{equations2}
    \end{align}
where $P=p+b^2/2$ is the total pressure, $d_i$ the ion inertial length and ${\bf f}$ a stationary homogeneous external force acting at large scales. 
The dissipation terms are
   \begin{align} 
{\bf d}_\nu =& \nu\Delta\vh \, , \\
{\bf d}_\eta =& \eta\Delta\va \, , 
    \end{align}
where $\nu$ is the kinematic viscosity and $\eta$ the magnetic diffusivity. For this system the equation of total energy conservation writes \citep{GCUP}
\begin{align} \label{consE}
   \partial_{t} \langle E^{tot} \rangle = \langle \vh \cdot {\bf d}_\nu \rangle + \langle \va \cdot {\bf d}_\eta \rangle + \langle \vh \cdot {\bf f} \rangle \, ,
\end{align}
where $\langle \rangle$ is an ensemble average, which is equivalent to a spatial average in homogeneous turbulence. 
We define the mean rate of total energy injection as $\varepsilon = \langle \vh \cdot {\bf f} \rangle$. 
With this, we can conclude that in the stationary regime the following relation holds
$\langle \vh \cdot {\bf d}_\nu + \va \cdot {\bf d}_\eta \rangle=-\varepsilon$.
Note that using the relation $\lang\textbf{X}\cdot\Delta\textbf{X}\rang =  - \lang(\nab \times \textbf{X})^{2}\rang$, which is valid for any incompressible vector field \textbf{X}, we also have:
\begin{align}
\langle  \vh \cdot {\bf d}_\nu\rangle + \langle \va \cdot {\bf d}_\eta \rangle = - \nu \left\langle {\bf w}^2 \right\rangle - \eta  \left\langle {\bf j}^2 \right\rangle \, , 
\end{align}
with ${\bf w} = \nabla \times \vh$ the vorticity, which gives the expression of the mean rate of total energy dissipation.

Next, we consider a spatial increment $\el$ connecting two points in space \textbf{x} and $\textbf{x}'$, as $\textbf{x}'=\textbf{x}+\el$, and we define $\vh \equiv \vh(\textbf{x})$ and $\vh' \equiv  \vh(\textbf{x}')$; the same notation is used for other variables. We also define the velocity increment $\delta\vh \equiv \vh'-\vh$. We recall that under this formalism, 
we have for any entity $A$:  $\partial_x A' = \partial_{x'} A = 0$. 
We then search for a dynamical equation for expression (\ref{correlator2}), under the hypothesis of statistical homogeneity, which means that we have to calculate 
$\partial_t \lang S\rang$. 
%
Using Eqs.~(\ref{equations0})--(\ref{equations2}) and the incompressibility of the flow we obtain,
    \begin{widetext}
     \begin{align} \label{v2_1}
         \centering    
        \partial_t  \vh^2  =&  -\nab\cdot[(\vh\cdot\vh)\vh] + 2\vh\cdot(\va\cdot\nab)\va - 2\vh\cdot\nab P + 2\vh\cdot {\bf d}_\nu + 2\vh\cdot {\bf f} \, , \\
   \partial_t  \va^2  =&  -\nab\cdot[(\va\cdot\va)\vh] + 2\va\cdot(\va\cdot\nab)\vh + d_i\nab\cdot[(\va\cdot\va)\ja] - 2d_i\va(\va\cdot\nab)\ja + 2\va\cdot {\bf d}_\eta \, , \label{v2_2} \\ 
        \partial_t (\vh\cdot\vh') =&  -\nab'\cdot[(\vh\cdot\vh')\vh'-(\vh\cdot\va')\va'+P'\vh] -\nab\cdot[(\vh\cdot\vh')\vh-(\vh'\cdot\va)\va+P\vh']  \nonumber \\
        &+ \vh' \cdot {\bf d}_\nu + \vh \cdot {\bf d'}_\nu + \vh' \cdot {\bf f} + \vh \cdot {\bf f'} \, , \label{v2_3} \\
        \partial_t  (\va\cdot\va')  =&  -\nab'\cdot[(\va\cdot\va')\vh' - (\va\cdot\vh')\va' - d_i(\va\cdot\va')\ja' + d_i(\va\cdot\ja')\va'] \nonumber \\ \label{v2_4}
        &- \nab\cdot[(\va'\cdot\va)\vh - (\va'\cdot\vh)\va - d_i(\va'\cdot\va)\ja + d_i(\va'\cdot\ja)\va] +  \va' \cdot {\bf d}_\eta + \va \cdot {\bf d'}_\eta \, , 
    \end{align}
    \end{widetext}
and similar equations as Eqs.~(\ref{v2_1})--(\ref{v2_2}) for the primed expressions. Below we will consider the ensemble average of the previous equations. 
We can use the relation $\lang\nab'\cdot\rang=-\lang\nab\cdot\rang=\nab_{\el} \cdot\lang\rang$, where $\nab_{\el}$ denotes the derivative operator along the increment $\el$, to suppress the pressure terms in Eqs.~(\ref{v2_1}) and (\ref{v2_3}),
    \begin{align} \nonumber   
       & \lang\vh\cdot\nab P\rang = \lang\nab\cdot(P\vh)\rang = -\lang\nab'\cdot(P\vh)\rang = 0 \, , \\ \nonumber
       & \lang \nab'\cdot(P'\vh)\rang = - \lang\nab\cdot(P'\vh)\rang = 0 \, . 
    \end{align}
By remarking that,
    \begin{align} \nonumber   
       \lang \vh\cdot(\va\cdot\nab)\va \rang =& - \lang \va\cdot(\va\cdot\nab)\vh \rang, \\ \nonumber
       \lang \va\cdot(\va\cdot\nab)\ja \rang =& - \lang \ja\cdot(\va\cdot\nab)\va \rang,
    \end{align}
a combination of Eq.~(\ref{v2_1}) to (\ref{v2_4}) leads to,
    \begin{widetext}
    \begin{align} \label{dynamic}
       \partial_t \lang S\rang =
       \, & 2 \nab_{\el} \cdot \lang (\vh\cdot\vh')\delta \vh + (\va\cdot\va')\delta \vh - (\vh\cdot\va') \delta \va - (\va\cdot\vh') \delta \va \rang \\ 
       &+ 2 d_i\nab_{\el} \cdot \lang - (\va\cdot\va')\delta \ja + (\va\cdot\ja')\va' - (\va'\cdot\ja)\va \rang 
       + 2 d_i\lang \ja\cdot(\va\cdot\nab)\va + \ja'\cdot(\va'\cdot\nab')\va' \rang \nonumber   \\ 
       &+ 4 \lang \vh \cdot {\bf d}_\nu \rang - 2\lang \vh \cdot {\bf d'}_\nu \rang - 2\lang \vh' \cdot {\bf d}_\nu \rang 
       + 4 \lang \va \cdot {\bf d}_\eta \rang - 2\lang \va \cdot {\bf d'}_\eta \rang - 2\lang \va' \cdot {\bf d}_\eta \rang
       + 4 \langle \vh \cdot {\bf f} \rangle - 2 \langle \vh \cdot {\bf f'} \rangle - 2 \langle \vh' \cdot {\bf f} \rangle \, . \nonumber
    \end{align}
    \end{widetext}
To further simplify expression (\ref{dynamic}) we can use the equalities $\nab\cdot[(\va\cdot\ja')\va] = \ja'\cdot(\va\cdot\nab)\va$ and 
$\nab'\cdot[(\va'\cdot\ja)\va'] = \ja\cdot(\va'\cdot\nab')\va'$ and relation (\ref{consE}). We then obtain,  
\begin{widetext}
\begin{align}    \label{compatible}
       \partial_t \lang S\rang =& - \nab_{\el} \cdot \lang (\delta\vh\cdot\delta\vh + \delta\va\cdot\delta\va)\delta\vh - 2 (\delta\vh\cdot\delta\va)\delta\va 
       - d_i (\delta\va\cdot\delta\va)\delta\ja + 2d_i (\delta\va\cdot\delta\ja)\delta\va \rang + 2d_i\lang \delta\ja\cdot\delta[(\va\cdot\nab)\va] \rang   \\
       & + 4 \partial_{t} \langle E^{tot} \rangle
        - 2\lang \vh \cdot {\bf d'}_\nu \rang - 2\lang \vh' \cdot {\bf d}_\nu \rang - 2\lang \va \cdot {\bf d'}_\eta \rang - 2\lang \va' \cdot {\bf d}_\eta \rang
       - 2 \langle \vh \cdot {\bf f'} \rangle - 2 \langle \vh' \cdot {\bf f} \rangle \, . \nonumber
\end{align}
\end{widetext}
It is interesting to note at this level that expression (\ref{compatible}) is fully compatible with the limit $\ell \to 0$ since in this case each term of the first line 
tends to $0$, and in the second line we have an exact compensation between the first term and the others by means of Eq.~(\ref{consE})

The final expression of the exact law for 3D IHMHD, valid in the inertial range, is obtained by using the stationarity assumption and the 
limit of a wide inertial range (i.e. large kinetic/magnetic Reynolds numbers limit) for which,
\begin{align}    
\lang \vh \cdot {\bf d'}_\nu \rang \simeq \lang \vh' \cdot {\bf d}_\nu \rang \simeq \lang \va \cdot {\bf d'}_\eta \rang \simeq \lang \va' \cdot {\bf d}_\eta \rang \simeq 0 \, , 
\end{align}
and also (with the properties of the external force) 
\begin{align}    
\lang \vh \cdot {\bf f'} \rang \simeq \lang \vh' \cdot {\bf f} \rang \simeq \varepsilon \, . 
\end{align}
We find the expression,
\begin{widetext}
\begin{align}    
       - 4\varepsilon =& \nab_{\el} \cdot \lang (\delta\vh\cdot\delta\vh + \delta\va\cdot\delta\va)\delta\vh - 2 (\delta\vh\cdot\delta\va)\delta\va 
       - d_i (\delta\va\cdot\delta\va)\delta\ja + 2d_i (\delta\va\cdot\delta\ja)\delta\va \rang - 2d_i\lang \delta\ja\cdot\delta[(\va\cdot\nab)\va] \rang \, .
\end{align}
\end{widetext}
This law can be written in a compact form as,
    \begin{align} \label{vKH}
       -4\varepsilon = \nab_{\el}\cdot (\textbf{Y}+\textbf{H}) - 2A \, , 
    \end{align}
    where
    \begin{align} \label{Y}
       \textbf{Y} &= \lang (\delta\vh\cdot\delta\vh + \delta\va\cdot\delta\va)\delta\vh -2(\delta\vh\cdot\delta\va)\delta\va \rang, \\\label{H}
       \textbf{H} &= d_i\lang 2(\delta\va\cdot\delta\ja)\delta\va - (\delta\va\cdot\delta\va)\delta\ja  \rang, \\ \label{A}
       A &= d_i\lang \delta\ja\cdot\delta[(\va\cdot\nab)\va] \rang.
    \end{align}
Here the contribution of the Hall effect is split into a flux \textbf{H} and a corrective term $A$. In the limit $d_i \to 0$ we recover the classic MHD law of \cite{PP98}. Note that Eq.~(\ref{vKH}) is the same as the one proposed in \cite{hellinger18} except for the corrective term $A$ which is here multiplied by a -2 factor (instead of 1). Assuming isotropy we can integrate expression (\ref{vKH}) which leads to, 
    \begin{align} \label{cascade}
       -\frac{4}{3}\varepsilon\ell = Y_{\el} + H_{\el} - 2I_A \, ,
    \end{align}
where $Y_\ell$ and $H_\ell$ are the projections along the displacement direction $\boldsymbol\ell$, respectively, and $I_A = (1/\ell^{2}) \int_{0}^{\ell}r^2Adr$.

Because the corrective term $A$ can prove to be difficult to compute in spacecraft data due to the term $\delta[(\va\cdot\nab)\va]$, we will see in the next section that H18 law can be improved and written using a simpler and more compact formulation involving only the $\textbf{H}$ term.

 \section{Alternative formulation of the corrected H18}

		To improve the corrected H18 law we need to do some calculation on the term $A$. Using the fact that $\nab(\textbf{X}\cdot\textbf{X}) = 2\textbf{X}\times(\nab\times\textbf{X}) + 2(\textbf{X}\cdot\nab)\textbf{X}$ and following a logic of calculation similar to \cite{BK18}, we have:
         \begin{align}
           A &= d_i\lang \delta\ja\cdot\delta[\frac{1}{2}\nab(\va\cdot\va)+\ja\times\va] \rang,
        \end{align} 
        and, using derivative properties along with $\nab\cdot\ja=0$, this equation reduces to: 
		\begin{align} \label{Awithdelta}
           A &= d_i\lang \delta\ja\cdot\delta(\ja\times\va) \rang.
        \end{align}
        Now, with the relation $\nab\cdot(\textbf{X}\times\textbf{Y}) = \textbf{Y}\cdot(\nab\times\textbf{X}) - \textbf{X}\cdot(\nab\times\textbf{Y})$ we can write (following \citep{banerjee17}),
        \begin{align} \nonumber
           \lang (\ja\times\va)\cdot\ja' \rang &= \lang (\ja\times\va)\cdot(\nab'\times\va') \rang \\\nonumber
           &= -\lang \nab'\cdot[(\ja\times\va)\times\va'] \rang  \\
           &= -\nab_{\el}\cdot\lang (\ja\times\va)\times\va' \rang \, , \\
           \lang (\ja'\times\va')\cdot\ja \rang &= \nab_{\el}\cdot\lang (\ja'\times\va')\times\va \rang, \\
           \lang (\ja'\times\va')\cdot\ja' \rang &= \lang (\ja\times\va)\cdot\ja \rang = 0,
        \end{align}
        which leads to,
        \begin{align} \label{Anewform}
          A &= d_i\nab_{\el}\cdot\lang (\ja \times \va)\times\va' - (\ja' \times \va')\times\va \rang.
        \end{align}
        Using identities for double cross product Eq.~(\ref{Anewform}) can be cast as,
        \begin{align} \nonumber
          A &= \frac{1}{2}d_i\nab_{\el}\cdot\lang 2(\delta\va\cdot\delta\ja)\delta\va - (\delta\va\cdot\delta\va)\delta\ja  \rang \\\nonumber
          &- d_{i}\nab_{\el}\cdot\lang (\va\cdot\ja')\va - (\va'\cdot\ja)\va' \rang \\\nonumber
          &= \frac{1}{2}\nab_{\el}\cdot \textbf{H} \\\nonumber
          &+ d_i\lang \ja'\cdot[(\va\cdot\nab)\va] + \ja\cdot[(\va'\cdot\nab')\va'] \rang \\
          &= \frac{1}{2}\nab_{\el}\cdot \textbf{H} - A,
        \end{align}
        and we obtain,
        \begin{align} \label{HtoA}
            \nab_{\el} \cdot \textbf{H} &= 4A \, .
        \end{align}
        Injecting relation (\ref{HtoA}) into expression (\ref{vKH}) we finally obtain the new formulation,
        \begin{align} \label{vKH2}
            -4\varepsilon = \nab_{\el}\cdot \left( \textbf{Y} + \frac{1}{2}\textbf{H} \right) \, , 
         \end{align}
which can be reduced to the following expression in the isotropic case,
        \begin{align} \label{F19}
            -\frac{4}{3}\varepsilon\ell = Y_{\el} + \frac{1}{2}H_{\el} \, . 
        \end{align}

This new formulation, which will be referred to as F19 hereafter, is one of the main results of this paper. It has the double advantage of depending only on the product of increments of the physical fields (unlike the G08 model) and of being expressed only as flux terms. This makes it easier to apply in particular to single spacecraft data (under the assumption of isotropy).

Below we will verify whether the law (\ref{vKH2}) derived above is compatible with the other IHMHD laws (G08 and BG17) in the inertial range. The testing will be focused on the Hall-induced terms \textbf{H} and $A$, as the ideal MHD term \textbf{Y} is exactly the same as the one from \cite{PP98b} and, by extension, the same as the ideal MHD component of G08. We will first investigate this question at the mathematical level and then with DNSs of EMHD. 

\section{Equivalence of the exact laws} \label{equiv}

    \subsection{Compatibility between F19 and G08}
    
        Here we show the equivalence of F19 and G08 by keeping only the Hall contributions. In G08, the law reads with our notation,
        \begin{align} \label{G08}
            -4\varepsilon_{Hall} &= 4 d_{i} \nab_{\el}\cdot\lang (\ja \times \va)\times\delta\va \rang.
        \end{align}
        We already showed that $\nab_{\el} \cdot \textbf{H} = 4A$. With Eq.~(\ref{Anewform}) we have,
        \begin{align} \nonumber
          \frac{1}{2}\nab_{\el}\cdot \textbf{H} &= 2d_i\nab_{\el}\cdot\lang (\ja \times \va)\times\va' - (\ja' \times \va')\times\va \rang \\
          &= 4d_i\nab_{\el}\cdot\lang (\ja \times \va)\times\va' \rang,
        \end{align}
        which is enough to show that,
		\begin{align} \label{hellVS2008}
           \frac{1}{2}\nab_{\el}\cdot\textbf{H} &= 4d_{i}\nab_{\el}\cdot\lang (\ja \times \va)\times\delta\va \rang \, , 
        \end{align}
        proving the compatibility.
        
\subsection{Compatibility between G08 and BG17}
    
Demonstrating the equivalence between the Hall terms of G08 and BG17 is even simpler. In the latter the Hall term is written, 
        \begin{align} \label{BG17}
            -4\varepsilon_{Hall} &= 2d_{i}\lang \delta(\ja\times\va)\cdot\delta\ja \rang \, .
        \end{align} 
		Using Eqs.~(\ref{Awithdelta}) and (\ref{HtoA}) we immediately obtain,
        \begin{align} \label{hellVS2017}
           \frac{1}{2}\nab_{\el}\cdot\textbf{H} &= \lang \delta(\ja\times\va)\cdot\delta\ja \rang,
        \end{align}
		and thus prove the equivalence.

\section{Numerical study}\label{num}

 \subsection{The equivalence of the models G08, BG17, F19}   
In this section we will compare the G08, BG17 and F19 laws by using 3D DNSs of incompressible EMHD turbulence (Eqs. (\ref{equations}) with $\textbf{v}=0$). We used a modified version of the TURBO code \citep{Teaca09} in which we have implemented the Hall effect \citep{meyrand13}.  The EMHD equations are solved in a triply periodic box. A pseudo-spectral algorithm is used to perform the spatial discretization on a grid with a resolution of $512^3$ collocation points (see \cite{meyrand13} for further details). A mean guide field $\va_0$ of magnitude unity is introduced along the z-axis. A large-scale forcing is applied which enforces a constant rate of energy injection with no helicity. The system is evolved until a stationary state is reached such that $\textbf{b}_{rms}\sim\va_0$. We removed the amount of ideal invariants that is injected into the system by the forcing mechanism by means of magnetic hyperdiffusivity $\eta_{3}\Delta^3$ with $\eta_{3}=10e^{-11}$. The data consists of three periodic cubes giving the three components of the magnetic field in each grid point. 
The values of $\varepsilon_{Hall}$ are obtained by averaging the mixed field increments of the different exact laws over all the points of the data cubes and spherically integrating them, using for the increment vectors $\el$ a set of specific directions in space defined by 73 base vectors as described in \cite{taylor03}, and lengths going from a three points distance to half the size of the cubes \citep[see also,][]{A2018b}.  

    \begin{figure}
        \centering
        \includegraphics[width=\hsize]{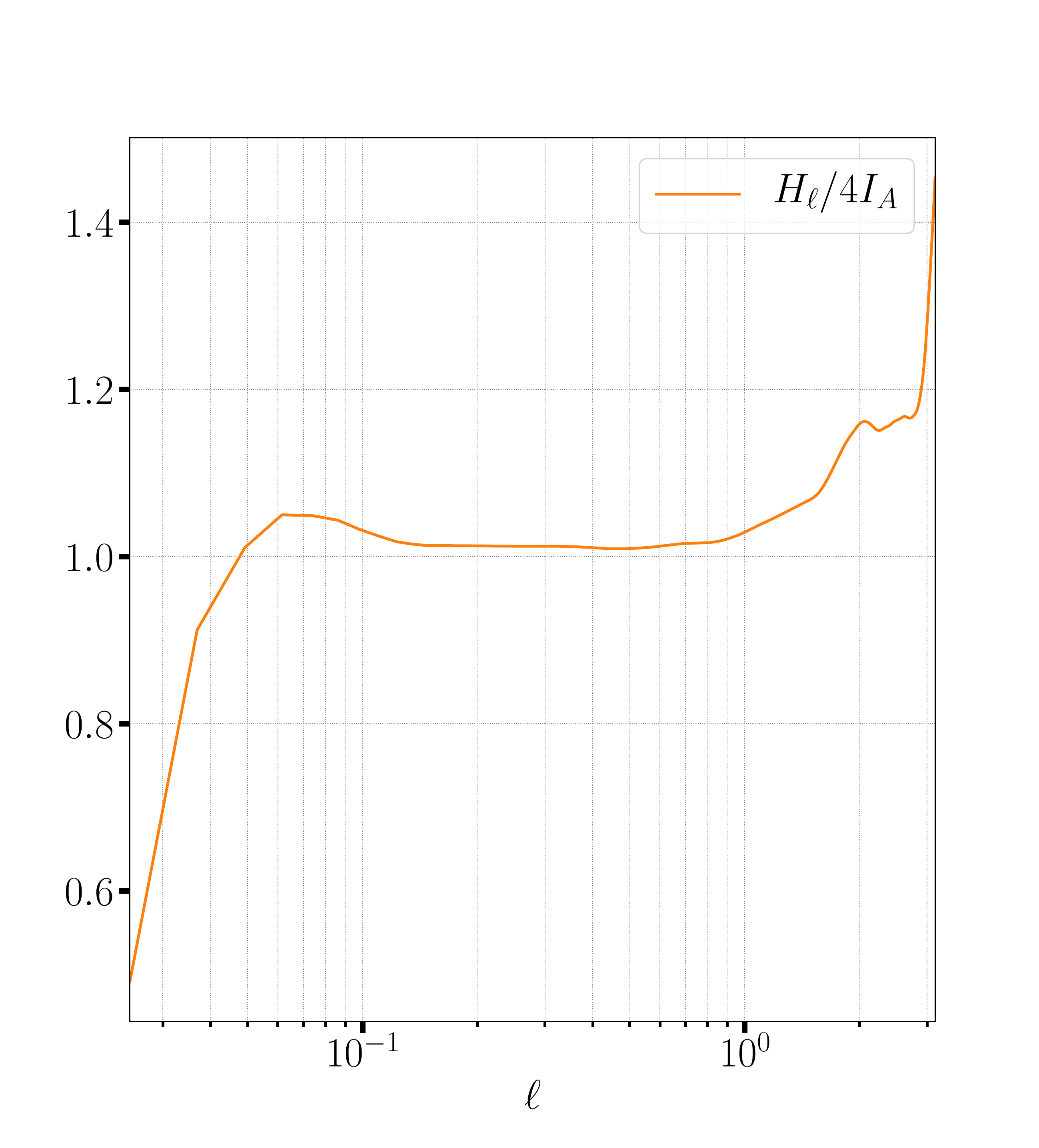}
        \caption{Ratio of $H_{\ell}$ to $4I_A$ of the F19 law.}
        \label{correction}
    \end{figure}
First of all, we want to check numerically the new law F19 and more precisely the analytical relation found between $H_{\ell}$ and $4I_A$. In Fig. \ref{correction} we represent $H_{\ell}/4I_A$, which shows differences mainly at large and small scales but not at intermediate scales where the inertial range is supposed to be. The differences observed are probably a consequence of the different nature of these two terms, being respectively a flux and an integrated term. The methods involved in the calculation being different, we can expect some minor differences. These should not alter the estimation of the energy cascade rate which is measured in the inertial range, ie. for scales $\ell \le 0.3$ (see below).
 
\begin{figure}
        \centering
        \includegraphics[width=\hsize]{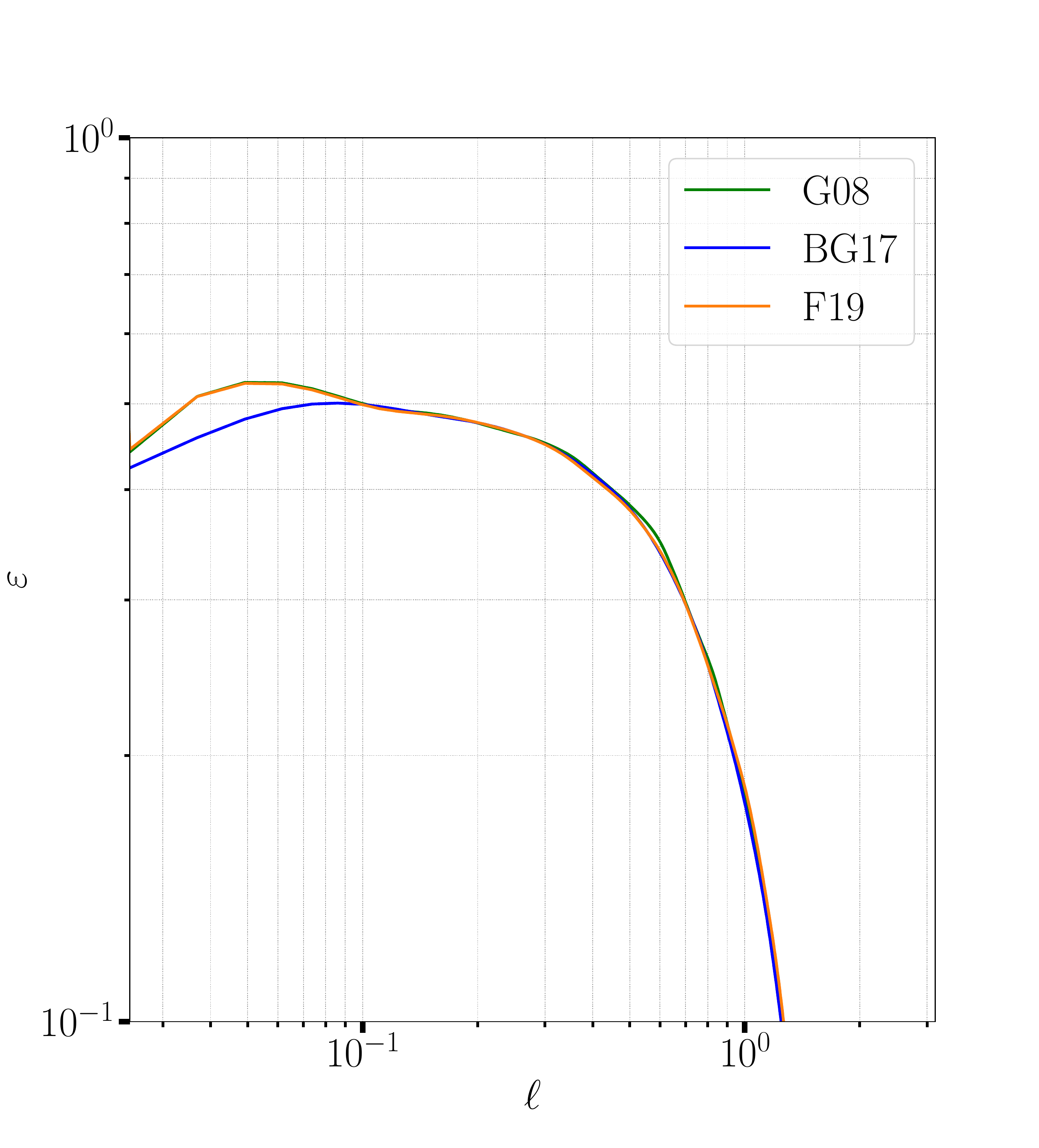}
        \caption{Energy cascade rates $\varepsilon_{Hall}$ computed with F19, $\text{G}08$ and BG17.}
        \label{second_comparison}
\end{figure}
   
To compare the three energy cascade rates obtained with the different expressions we first note that the Hall term in the G08 model can be written as
 \begin{align} \nonumber
           \text{G}08_{Hall} &= 2\nab_{\el}\cdot\lang (\ja \times \va)\times\delta\va + (\ja' \times \va')\times\delta\va \rang.
    \end{align}
This formulation is chosen (over other possible expressions) because it ensures a symmetry between \textbf{x} and $\textbf{x}'$ as in the two other laws, F19 and BG17. One must also be careful when computing BG17, as this law gives the energy cascade rate as a function of a direct statistical mean and not a flux, and thus does not require an integration \textit{a priori}. However, we need to keep in mind that $\varepsilon$ is not, in fact, exactly constant in our data. Consequently, when we integrate F19 and G08, we compute in reality $(1/\ell^{3}) \int_{0}^{\ell}r^2\varepsilon dr$ and not $\varepsilon$. To remain consistent between the three models, we need to use the non integrated forms of both F19 and G08 (\ref{vKH2},\ref{G08}). This is what will be done hereafter.

We computed the energy cascade rate from the three laws and obtained the results gathered in Fig. \ref{second_comparison}. All three laws fit remarkably well with each other, with however a slightly different behaviour of BG17 model at scales $\ell \le 0.1$. Using the non integrated forms of G08 and F19 required us to apply a discrete derivation to our results as we only compute the inner bracket of the flux terms, and we expect this operation to be responsible for the differences at small scales due to a lack of resolution in this range of scales. The inertial range induced by the Hall effect is not easy to pinpoint precisely, but can be roughly estimated as going from 0.05 to 0.3 in this simulation.
 \subsection{On the role of $\va_0$}   
Finally, we tested the influence of a guide field $\va_0$ on the estimation of $\varepsilon_{Hall}$. Indeed, the introduction of a uniform magnetic field $\va_0$ into the previous laws does not change their expression. This is obvious for F19 which only depends on increments and for BG17 in which the $\va_0$ influence translates as $\lang \delta\ja\cdot((\delta\ja)\times\va_0) \rang = 0$. For G08 we have,
\begin{align} \nonumber
	\text{G}08_{Hall} &= 2\nab_{\el}\cdot\lang (\ja \times \va_0)\times\va' - (\ja' \times \va_0)\times\va \rang \\\nonumber
	&= -2 \lang (\ja \times \va_0)\cdot\ja' + (\ja' \times \va_0)\cdot\ja \rang \\
	&= 0 \, .
\end{align}
Thus, when computing $\epsilon$ taking or not $\va_0$ into account in the data should not affect the result. However, for pratical reasons related to the numerical compuation, $\va_0$ may have some influence on estimating $\epsilon$ as we show now.

\begin{figure}
        \centering
        \includegraphics[width=\hsize]{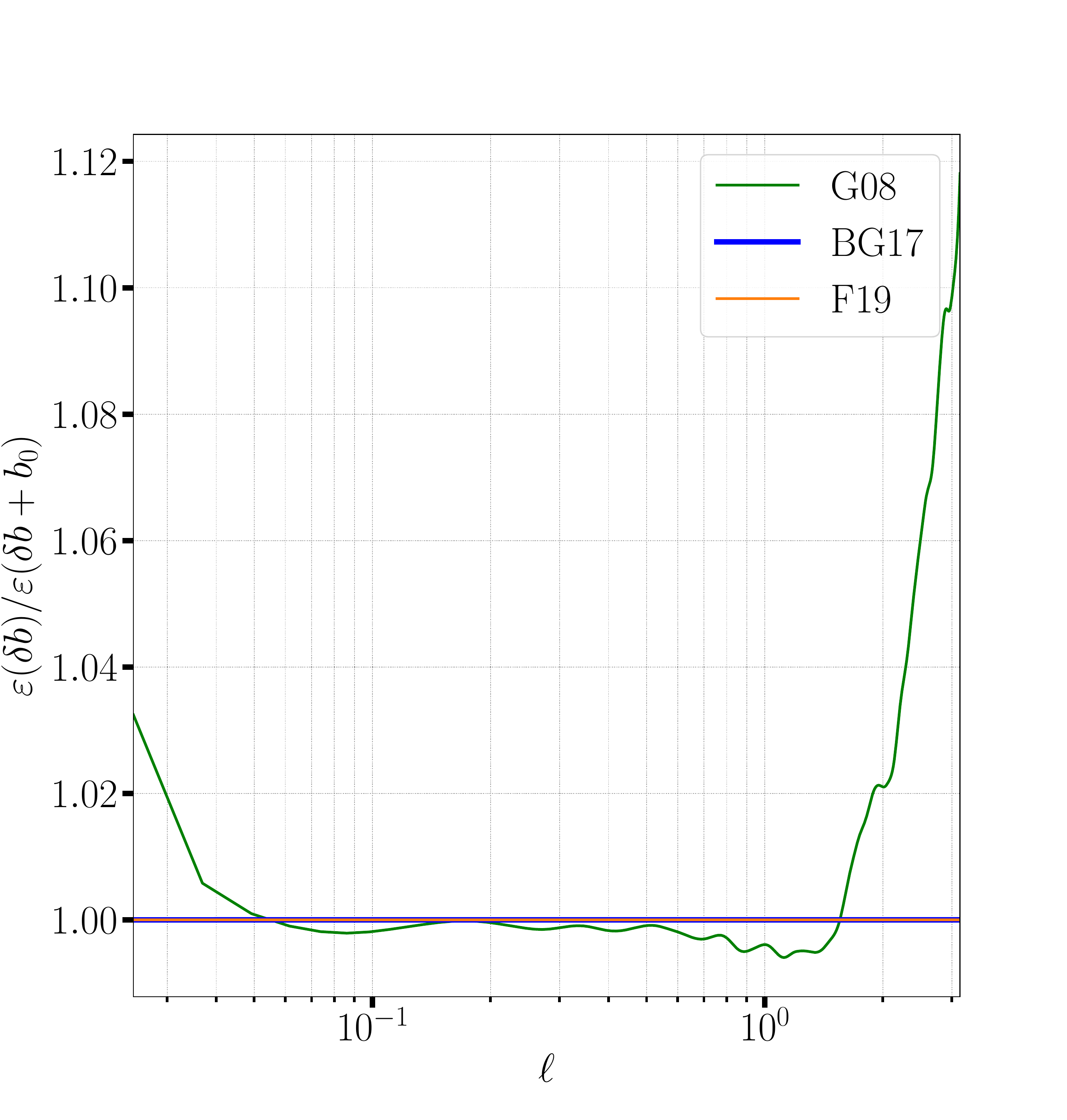}
        \caption{Ratios of $\varepsilon_{Hall}$ computed from the data cubes where the guide field $\va_{0}$ is removed and from the data cubes where it is not. The values obtained with F19 and BG17 overlap.}
        \label{field_impact}
    \end{figure}

Values of $\varepsilon_{Hall}$ computed only with the fluctuating magnetic fields were obtained by averaging the magnetic field component along the guide field axis (here the z axis) and subtracting this value from that component. In Fig. \ref{field_impact} we see that computing the energy cascade rate with or without the mean guide field leads to the same result for all but G08, even though the contribution of $\va_0$ reduces to zero mathematically. The difference is, however, very small in the inertial range (less than 0.5\%).

We believe this problem to be tied to the way we handle derivatives. F19 is formed of only increments and so does not involves $\va_0$, unlike the models BG17 and $\text{G}08$ that contain a $\va_0$ contribution \textit{a priori}, but whose contributions in fact reduce to zero. However, from these contributions, only the one in $\text{G}08$ comes from a flux term and so is preceded by a derivative. When we compute numerically the energy cascade rates we do not really calculate this derivative but rather use the approximation $\nab_{\el} \rightarrow 1/\ell$. Thus, we are making an approximation in the calculation and this may be the cause of the behavior shown in Fig. \ref{field_impact}. A similar remark was made in \cite{hadid17} which led us to this conclusion. It may also be worth to mention that the validity of this approximation is tied to the validity of the hypotheses of isotropy and homogeneity, and the influence of $\va_0$ would probably be more important when using observational data where these hypotheses are harder to meet. 

Based on these remarks and on the behavior of the three laws we conclude that, as expected, $\va_0$ does not contribute explicitly to the incompressible energy cascade rate and, in the purpose of computing $\varepsilon_{Hall}$, that it should be removed from the simulation of the spacecraft data beforehand in order to minimize the possible numerical errors that it can generate. Note that this property does not mean that $\va_0$ has no influence on the nonlinear dynamics \citep{galtier2000,W2012,Oughton13}: it is actually expected that the energy cascade rate $\varepsilon_{Hall}$ decreases with increasing $b_0$, as shown recently with DNSs \citep{Bandy18}.
It is worth mentioning that the situation is very different in compressible law \citep[e.g.,][]{Banerjee13} where the $\va_0$ dependence is explicit and cannot {\it a priori} be ruled out~\citep{hadid17}. However, recent development suggest that its influence will not be significant as it mostly impact the volumic contibutions to the law, which appear to be small compared to the dominant flux terms \citep{A2018b}.

\section{Conclusion}
\label{conclusion}

    The energy cascade rate $\varepsilon$ is an essential tool for studying turbulent flows. Despite being sometimes hard to compute it can be theoretically calculated by several yet equivalent formulations. We showed here that the law (\ref{vKH}), which is obtained using the same premises as proposed in \cite{hellinger18}, can be written (when corrected) in a compact form with only a flux term (\ref{vKH2}). As shown numerically, this gives the same energy cascade rate in the inertial range as with the G08 and BG17 laws. This diversity of exact laws gives more freedom to compute the energy cascade rate of IHMHD turbulence as it is possible to adapt the computation method to the data available and their quality.
    
   For instance, we showed that the presence of a mean guide field should not contribute explicitly to the energy cascade rate. This theoretical property is well verified with a DNSs for BG17 and F19, but not for G08 which shows a dependence on $\va_0$ that can be interpreted as residual errors due to the performed computation. Although this dependence remains small in the present paper, it is more important in spacecraft data analysis \citep{private}. Therefore we advise using F19 or BG17 laws to compute the energy cascade rate as they are free from the errors induced by the presence of a mean guide field.


\section*{Acknowledgements} 
The authors acknowledge financial support from CNRS/CONICET Laboratoire International Associ\'e (LIA) MAGNETO. N.A. is supported through a DIM-ACAV post-doctoral fellowship. S. B. acknowledges DST Inspire research grant.

\bibliography{Ref_final.bib}
\end{document}